\documentclass[conference]{IEEEtran}
\IEEEoverridecommandlockouts

\usepackage{cite}
\usepackage{amsmath,amssymb,amsfonts}
\usepackage{algorithmic}
\usepackage[dvipdfmx]{graphicx}
\usepackage{textcomp}
\usepackage{xcolor}
\def\BibTeX{{\rm B\kern-.05em{\sc i\kern-.025em b}\kern-.08em
    T\kern-.1667em\lower.7ex\hbox{E}\kern-.125emX}}

\usepackage{latexsym}
\usepackage{url}

\newtheorem{definition}{Definition}

\newcommand{\Setup}{{\sf Setup}}
\newcommand{\OKGen}{{\sf OKGen}}
\newcommand{\UKGen}{{\sf UKGen}}
\newcommand{\RSign}{{\sf RSign}}
\newcommand{\RVerify}{{\sf RVerify}}
\newcommand{\Open}{{\sf Open}}
\newcommand{\Judge}{{\sf Judge}}
\newcommand{\pp}{\mathsf{pp}}
\newcommand{\opk}{\mathsf{opk}}
\newcommand{\osk}{\mathsf{osk}}
\newcommand{\pk}{\mathsf{pk}}
\newcommand{\sk}{\mathsf{sk}}

\newcommand{\vk}{\mathsf{vk}}
\newcommand{\sigk}{\mathsf{sigk}}

\newcommand{\Policy}{\mathsf{Policy}}
\newcommand{\DeployContractWallet}{\mathsf{DeployContractWallet}}
\newcommand{\SendTransaction}{\mathsf{SendTransaction}}
\newcommand{\Code}{\mathsf{Code}}
\newcommand{\BlockChain}{\mathsf{BlockChain}}
\newcommand{\TransactionData}{\mathsf{TransactionData}}
\newcommand{\CW}{\mathsf{CW}}

\begin{document}

\title{An Anonymous yet Accountable Contract Wallet System using Account Abstraction
}

\author{
\IEEEauthorblockN{Kota Chin}
\IEEEauthorblockA{\textit{University of Tsukuba}\\\textit{National Institute of}\\\textit{Information and}\\\textit{Communications Technology}\\\textit{Japan}}\\
\and
\IEEEauthorblockN{Keita Emura}
\IEEEauthorblockA{\textit{Kanazawa University, Japan}\\\textit{National Institute of}\\\textit{Information and}\\\textit{Communications Technology}\\\textit{Japan}}\\
\and 
\IEEEauthorblockN{Kazumasa Omote}
\IEEEauthorblockA{\textit{University of Tsukuba}\\\textit{National Institute of}\\\textit{Information and}\\\textit{Communications Technology}\\\textit{Japan}}\\
}

\maketitle

\begin{abstract}
Account abstraction allows a contract wallet to initiate transaction execution. 
Thus, account abstraction is useful for preserving the privacy of externally owned accounts (EOAs) because it can remove a transaction issued from an EOA to the contract wallet and hides who issued the transaction by additionally employing anonymous authentication procedures such as ring signatures. 
However, unconditional anonymity is undesirable in practice because it prevents to reveal who is accountable for a problem when it arises. 
Thus, maintaining a balancing between anonymity and accountability is important. In this paper, we propose an anonymous yet accountable contract wallet system. In addition to account abstraction, the proposed system also utilizes accountable ring signatures (Bootle et al., ESORICS 2015). The proposed system provides (1) anonymity of a transaction issuer that hides who agreed with running the contract wallet, and (2) accountability of the issuer, which allows the issuer to prove they agreed with running the contract wallet. Moreover, due to a security requirement of accountable ring signatures, the transaction issuer cannot claim that someone else issued the transaction. This functionality allows us to clarify the accountability involved in issuing a transaction. 
In addition, the proposed system allows an issuer to employ a typical signature scheme, e.g., ECDSA, together with the ring signature scheme. This functionality can be considered an extension of the common multi-signatures that require a certain number of ECDSA signatures to run a contract wallet. 
The proposed system was implemented using zkSync (Solidity). 
We discuss several potential applications of the proposed system, i.e., medical information sharing and asset management.
\end{abstract}


\section{Introduction}

\subsection{Introduction of Account Abstraction}

Ethereum involves two kinds of accounts, i.e., externally owned accounts (EOA), which are controlled by a user-managed secret key, and contract accounts (contract wallets), which are controlled by smart contracts. In the current implementation of Ethereum, to run a contract wallet, an EOA must send a transaction to the contract wallet. Then the contract wallet runs a transaction according to the rule specified in the contract. 
Account abstraction~\cite{AA} allows a contract wallet to initiate transaction execution, i.e., it can remove a transaction sent from an issuer to the contract.
Account abstraction provides two primary benefits. The first benefit is the reduction of gas costs because account abstraction can remove a transaction sent from an issuer to the contract wallet. The second benefit is flexible verification. In the current Ethereum implementation, transactions, issued by EOAs, are verified according to the validity of signatures generated by secret keys of EOAs, and the underlying signature scheme is restricted to the elliptic curve digital signature algorithm (ECDSA). Thus, ECDSA signatures are employed generally when transaction validity is verified in the contract, although, theoretically, any signature schemes can be employed because signatures are verified by programs. 
In contrast, due to account abstraction, no EOA issues transactions, and thus any signature schemes can be employed more easily to verify transaction validity  in the contract. For example, CRYSTALS-Dilithium~\cite{DucasKLLSSS18}, FALCON~\cite{Falcon}, and SPHINCS+~\cite{SPHINCS+}, which have been selected by the NIST Post-Quantum Cryptography Standardization, can be employed under the assumption they can be implemented by a program language that can be run by the contract wallet, e.g., Solidity.%
\footnote{Although we consider only signatures in this paper, any verification method can be employed or a contract wallet with no verification step can be constructed.} In addition, account abstraction allows us to employ signatures with rich functionalities, such as accountable ring signatures~\cite{BootleCCGGP15}. 
Representative examples of systems that support account abstraction include StarkNet\footnote{\url{https://starkware.co/starknet/}} and zkSync\footnote{\url{https://zksync.io/}}, which are network technologies in Ethereum Layer 2 (L2). Note that L2 is described in further detail in Section~\ref{L2}.

Transactions between the EOAs and contract wallets are removed; thus account abstraction is useful for preserving the privacy of EOAs. In fact, EIP-2938~\cite{AA} states that \lq\lq \emph{Privacy-preserving systems like tornado.cash}"  are a motivation to introduce account abstraction (tornado.cash is a mixing service). Concretely, it is expected that account abstraction can hide the issuer of a transaction. Note that a contract wallet typically verifies a signature using a verification key that has been registered in the contract program. Since the verification key is public (in a public blockchain), anyone can identify who issued the transaction (precisely, which verification key is used and the key holder who issued the transaction). In other words, account abstraction does not hide transaction issuer information. We emphasize that unconditional anonymity could promote crime. 
For example, tornado.cash is a tool that could be misused to facilitate money laundering, and the Fiscal Information and Investigation Service, an agency of the government of the Netherlands, arrested a 29-year-old man in Amsterdam as a suspected developer.%
\footnote{Arrest of suspected developer of {T}ornado {C}ash: \url{https://www.fiod.nl/arrest-of-suspected-developer-of-tornado-cash/} (August 12, 2022)} 
This indicates maintaining a balancing between anonymity and accountability is important. 

\subsection{Privacy on Ethereum}
Each EOA manages an ECDSA verification key $\vk$, and the corresponding address is a (last 20 bytes of) hashed value of $\vk$ (here Keccak-256 hash function is used). Here, as each address appears random,%
\footnote{Vanity addresses are often used to reduce the storage cost of addresses. For example, using an address 0x0000..., a part of address \lq\lq 0$\cdots$0" does not need to be stored. Even in this case, pseudonymity level privacy protection is guaranteed. }
 anyone can easily determine whether multiple transactions were issued by the same EOA, although it is difficult to detect the EOA in the real world. This means that issuer privacy is preserved in the sense of pseudonymity in crypto asset trading.

\medskip
\noindent\textbf{Does Pseudonymity Provide Sufficient Privacy Protection?} When crypto assets are traded between individuals, the pseudonymity level privacy protection might be sufficient. However, some information leakage occurs if a system is implemented using smart contracts. 

For example, assume that a request to view patient data is made by a transaction in an electronic medical record system using Ethereum smart contracts~\cite{OmarBBKR19,8904712,7573685,GeorgeC22,FanWRLY18,7990130,ROEHRS201770,9233462,electronics10050580}. In this case, we assume that multi-patient data are accessed from the same address. Then, we expect that the transaction issuing address is managed by a doctor or researcher. In addition, we expect that patients are suffering from similar diseases. Here, if the doctor or researcher associated with the address is identifiable, information identifying the patient's disease may be leaked (although this was not authorized by the patient) depending on the specialty of the issuer of the transaction. 
Thus, no information about hospitals and medical offices should be leaked from transactions. However, it is necessary to internally verify who issued the transaction in order to prevent unnecessary access to medical data. 

In the case of group asset management, pseudonymity level privacy protection would make the investment performance of each address public. Thus, it is desirable to be able to prove the investment performance internally while keeping information about the issuer of the transaction secret externally.

\medskip
\noindent\textbf{A Naive Solution 1: Key Sharing}. Let a pair of signing and verification keys be generated and let all members of a group share the key pair. When a member issues a transaction, the same signing key is used to generate a signature. Then, anonymity level privacy protection is provided because a contract wallet uses the same verification key to verify the signature. However, there is no way to identify who issued the transaction, even among group members. Unconditional anonymity is undesirable in practice because it prevents to reveal who is accountable for a problem when it arises. 
In addition, determining how to revoke the signing key when a member leaves the group is a nontrivial problem. 

\medskip
\noindent\textbf{A Naive Solution 2: External Services}. Amazon Web Services (AWS) Key Management Service or a system on permissioned blockchain that support a trusted execution environment (TEE), e.g., Intel SGX, can be used to solve the above problems. 
By using such services, group members can issue a transaction without sharing the signing key. 
Here, anonymity still holds because the contract wallet uses the same verification key. 
In addition, group members can internally identify who issued a transaction via an access log (AWS Key Management Service) or the transaction (permissioned blockchain). Member revocation is also possible via access control to the services. 
Thus, technically we can provide both anonymity (for outside of the group) and accountability (for the group) simultaneously. 
However, both methods assume trust in AWS and TEE, which is undesirable relative to increase trust points. 

\begin{figure*}[t]
\centering
\includegraphics[clip,width=12cm]{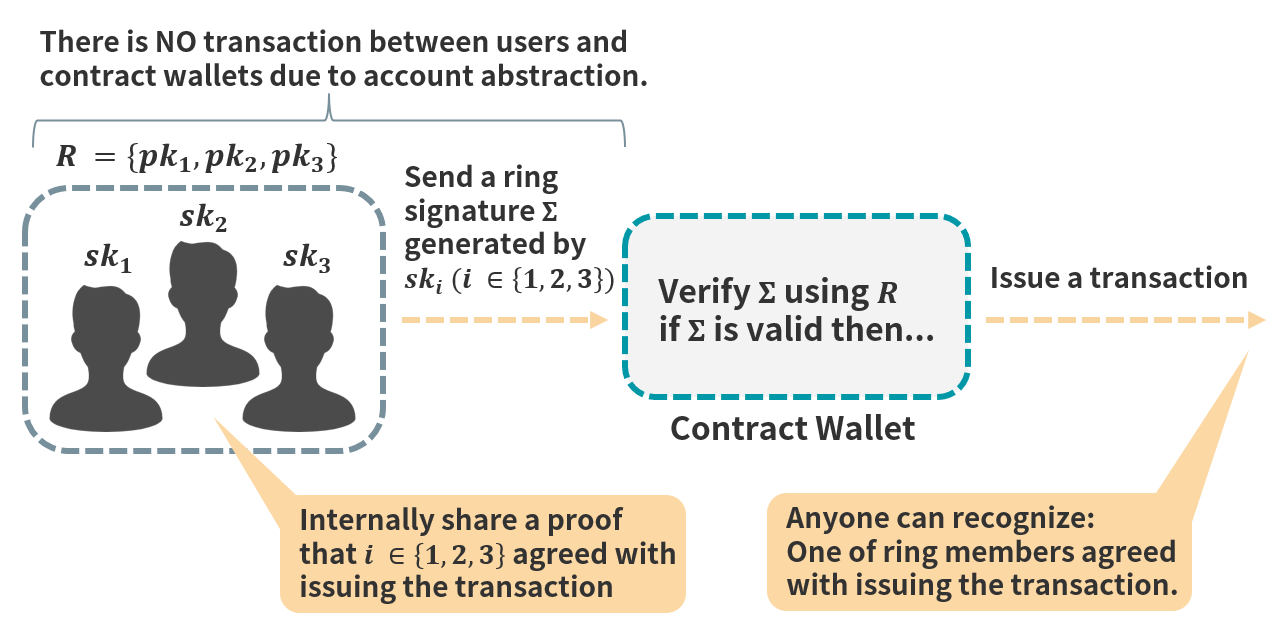}
\vspace{-0.3cm}
\caption{Proposed System}\label{OurSystem}
\end{figure*}

\subsection{Our Contribution}

In this paper, we focus on account abstraction, which is attractive in terms of constructing a privacy-preserving system, and we propose an anonymous yet accountable contract wallet system (Figure~\ref{OurSystem}).%
In addition to account abstraction, the proposed system employs accountable ring signatures~\cite{BootleCCGGP15}.%
\footnote{A user (EOA) needs to issue a transaction to a contract wallet when account abstraction is not employed. Thus, information that who issued a transaction is leaked even a ring signature scheme is employed in this case. 
Thus, both account abstraction and ring signatures are required mandatory in our system. We do not the following one-time deployment case that a contract wallet is newly deployed for issuing a transaction because it is not realistic solution.} We implemented the proposed system using zkSync (Solidity). 
Precisely, we implemented the verification algorithm of the underlying accountable ring signature scheme using Solidity. 
The proposed system is briefly introduced as follows. 

\begin{figure*}[t]
\centering
\includegraphics[clip,width=13cm]{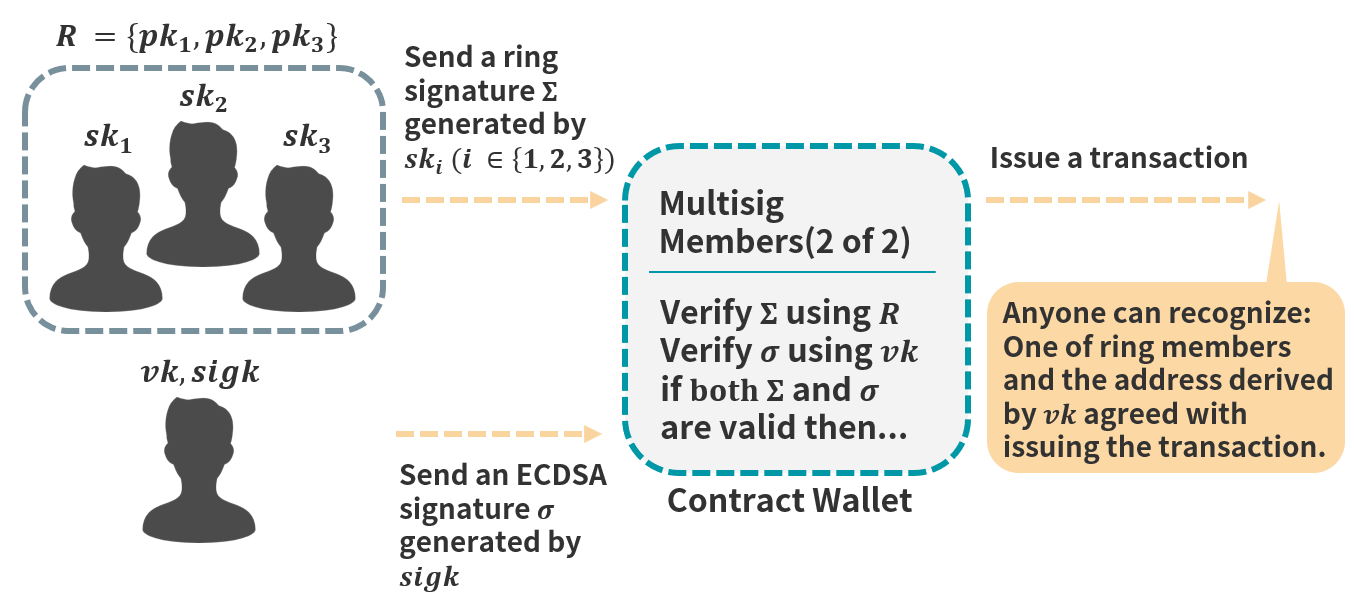}
\vspace{-0.3cm}
\caption{Extended Multi-Signatures}\label{ExtendedMultiSig}
\end{figure*}

\begin{itemize}
\item To issue a transaction, a ring signature is sent to a contract wallet, and the contract wallet verifies the signature. Then, the contract wallet uses a ring (i.e., a set of verification keys) to verify the signature. Anonymity holds in the sense that the contract wallet cannot identify the issuer among the verification key holders. 

\item The opening functionality of the accountable ring signatures allows the actual issuer to prove that they issued the transaction, and other users can recognize this fact. In contrast, due to a security requirement of accountable ring signatures (i.e., opening soundness), no user can prove that they issued a transaction if they did not actually issue the transaction, and moreover the transaction issuer cannot claim that someone else issued the transaction. In addition, the proof can be generated by other ring members who share the secret opening key of the underlying accountable ring signatures. This functionality allows us to clarify the accountability involved in issuing a transaction. 
We remark that we mainly focus on how to identify/prove who issued a transaction in this paper. That is, we assume that the organization to which the transaction issuer belongs is accountable, and we do not address what penalties that organization would impose on the transaction issuer if the transaction became problematic, or how that organization would be held accountable. 


\item The proposed system allows an issuer to employ a typical signature scheme, e.g., ECDSA, with the ring signature scheme. This functionality can be considered an extension of the typical multi-signatures that require a certain number of ECDSA signatures to run a contract wallet. For example, the contract wallet determines whether one of the issuers belongs to a group (in an anonymous manner), and a specific user (in the usual manner) agrees to run the contract wallet by verifying a ring signature and an ECDSA signature. This is illustrated in Figure~\ref{ExtendedMultiSig}. 
In the usual multi-signatures setting, anyone can recognize who agreed with running the contract wallet by observing ECDSA verification keys used for ECDSA signatures verification. 
Note that if two signatures are sent to the contract wallet separately, one of them must be preserved in the contract wallet which incurs an additional gas cost. To reduce the cost, one sender can collect all  of the required signatures, and send them to the contract wallet.  

\end{itemize}

\subsection{Applications}

Potential applications of the proposed system include, but are not limited to, the following.

\medskip
\noindent\textbf{Medical Information Sharing}. As mentioned previously, when a request to view patient data is made by a transaction in an electronic medical record system using Ethereum smart contracts, it is highly desirable that no information about hospitals and medical offices is leaked from the transactions. However, it is also essential to be able to internally verify who issued the transaction in order to prevent unnecessary access to medical data. 
With the proposed system, from the outside, it is possible to hide who is accessing patient data among members in an organization. For example, when a hospital is specified as the organization, it is possible to keep secret which doctors' offices have access to medical data. If a clinical department, e.g., internal medicine, psychosomatic medicine, or plastic surgery, is specified as the organizational unit, which department is related to which disease is leaked, but no other information is leaked because which doctor accessed the patient data is kept secret. In addition, dividing the organization into independent departments relative to the actual medical offices may further prevent unnecessary leakage of information. Thus, by setting organizational units appropriately, we can control leaked information. 
In addition, more flexible settings can be realized by using our extended multi-signatures. For example, we can consider a case where the hospital director or a department head must agree to issue a transaction. Here, one person in a medical office needs to agree to issue the transaction. 

\medskip
\noindent\textbf{Asset Management}. 
The range of blockchain-based asset management has expanded due to the advent of smart contracts. Thus, the number of investment companies that manage their clients' funds has been increasing. In some cases, the addresses held by lenders are known publicly. Then, the status of the fund management can be checked.%
\footnote{In fact, these addresses are not disclosed by lenders intentionally. Here, they are assumed to be lenders' addresses based on the asset management status. } It is assumed that there are cases where multiple users share an EOA as the address, and cases where multiple users employ contract wallets where transactions can be issued with the consent of a certain number of users. In the former case, it is impossible to know who issued the transaction internally. In the latter case, due to the pseudonymity level privacy protection, anyone can check the investment performance of each address. Except for cases where the investment status is disclosed intentionally, such information can be leaked unexpectedly. 
By employing the proposed system, the investment status of each address is not disclosed externally, and the investment status of each individual can be known internally. 

\section{Preliminaries}

\subsection{Accountable Ring Signatures}

As a generalization of ring signatures~\cite{RivestST01} and group signatures~\cite{ChaumH91}, Xu and Yung proposed accountable ring signatures~\cite{XuY04}. 
Briefly, as in ring signatures, each user generates their own public key $\pk$ and secret key $\sk$. This decentralized structure matches blockchain systems. 
An opener has a public key $\opk$ and a secret key $\osk$. When a user who has $(\pk,\sk)$ generates a ring signature $\Sigma$ on a message $M$, the user selects a set of public keys, which we refer to as ring $R$, and we assume that $\pk\in R$, and selects the opener by indicating $\opk$. 
We say that $(\Sigma,M)$ is valid if the signer is a member of $R$, i.e., there exists $\pk\in R$ for which the corresponding $\sk$ has been used to generate $\Sigma$. As in group signatures, the designated opener can trace the signer using $\osk$. Moreover, the opening algorithm produces a proof $\pi$ proving that $\Sigma$ is generated by $\sk$ corresponding to $\pk\in R$. 
We introduce the syntax defined by Bootle et al.~\cite{BootleCCGGP15} as follows. 

\medskip
\begin{definition}[Syntax of Accountable Ring Signatures~\cite{BootleCCGGP15}]~
\begin{itemize}
\item $\Setup(1^\lambda)$: The setup algorithm takes the security parameter $\lambda\in\mathbb{N}$ as input and outputs the common parameter $\pp$. 

\item $\OKGen(\pp)$: The opener key generation algorithm takes $\pp$ as input and outputs the opener public and secret keys $\opk$ and $\osk$, respectively. 

\item $\UKGen(\pp)$: The user key generation algorithm takes $\pp$ as input and outputs the user public verification key $\pk$ and user secret signing key $\sk$. 

\item $\RSign(\opk,M,R,\sk)$: The signing algorithm takes $\opk$, a message $M$  to be signed, a ring $R$, and $\sk$ as inputs and outputs a ring signature $\Sigma$. Here, $R$ is a set of user public keys and the $\pk$ corresponding to $\sk$ is assumed to be $\pk\in R$. 

\item $\RVerify(\opk,M,R,\Sigma)$: The verification algorithm takes $\opk$, $M$, $R$, and $\Sigma$ as inputs and outputs 1 (accept) or 0 (reject). 

\item $\Open(M,R,\Sigma,\osk)$: The open algorithm takes as $M$, $R$, $\Sigma$, and $\osk$ as inputs and outputs $\pk\in R$ of the signer and its proof $\pi$ or $\bot$ otherwise. 

\item $\Judge(\opk,M,R,\Sigma,\pk,\pi)$: The judge algorithm takes $\opk$, $M$, $R$, $\Sigma$, $\pk$, and $\pi$ as inputs and outputs 0 if $\RVerify(\opk,M,R,\Sigma)=0$; otherwise, it outputs 1 to indicate that $\Sigma$ is generated by $\sk$ corresponding to $\pk$, and 0 otherwise. 
\end{itemize}
\end{definition}

We require correctness holds where an honestly generated signature is always valid (the $\RVerify$ algorithm outputs 1), and a proof generated by the $\Open$ algorithm against the signature and corresponding verification key is always accepted by the $\Judge$ algorithm. 
Bootle et al. defined full unforgeability, anonymity, traceability, and tracing soundness, which are briefly explained as follows. Refer to the literature~\cite{BootleCCGGP15} for details on these security definitions. 

\begin{itemize}
\item Full Unforgeability: It ensures that no adversary $\mathcal{A}$ can falsely accuse an honest user of creating a ring signature, nor $\mathcal{A}$ can forge a ring signature on behalf of an honest ring. 
$\mathcal{A}$ takes $\pp$ as input and outputs $(\opk,\pk,M,R,\Sigma,\pi)$. Assume that $\mathcal{A}$ does not have $\sk$ corresponding to $\pk$ and $R$, and does not obtain a signature via the signing query $(\opk,\pk,M,R)$. $\mathcal{A}$ wins if $\RVerify(\opk,M,R,\Sigma)=1$ or $\Judge(\opk,M,R,\Sigma,\pk,\pi)=1$. Note that $\mathcal{A}$ can control the opener since $\mathcal{A}$ produces $\opk$. 

\item Anonymity: It ensures that no adversary $\mathcal{A}$ (who does not have $\osk$) can identify the signer, i.e., a signature does not reveal the identity of the ring member who generated it. 
$\mathcal{A}$ takes $(\pp,\opk)$ as input, declares two signing keys $(\sk_0,\sk_1)$, $M$ and $R$, and obtains $\Sigma^\ast\leftarrow \RSign(\opk,M,R,\sk_b)$ where $b\in\{0,1\}$ is the challenge bit. $\mathcal{A}$ outputs $b^\prime\in\{0,1\}$ and wins if $b=b^\prime$. Note that $\mathcal{A}$ knows all signing keys since $\mathcal{A}$ produces $(\sk_0,\sk_1)$. It guarantees that anonymity holds under the full key exposure. 

\item Traceability: It ensures that no adversary $\mathcal{A}$ can produce a signature that is valid but untraceable. 
$\mathcal{A}$ takes $\pp$ as input and outputs $(\osk,M,R,\Sigma)$. Assume that $\osk$ deterministically defines the corresponding $\opk$. Let $(\pk,\pi)\leftarrow \Open(M,R,\Sigma,\osk)$. $\mathcal{A}$ wins if $\RVerify(\opk,\allowbreak M,R,\Sigma)=1$ and $\Judge(\opk,\allowbreak M,R,\Sigma,\pk,\pi)=0$. Note that $\mathcal{A}$ can produce all secret keys. It guarantees that nobody, including a valid user who has $\sk$ and the opener who has $\osk$, can produce a valid but untraceable signature.

\item Tracing Soundness: It ensures that no adversary $\mathcal{A}$ can produce proofs for a signature that are accepted by the $\Judge$ algorithm for different verification keys. 
$\mathcal{A}$ takes $\pp$ as input and outputs $(M,\Sigma,R,\opk,\pk_1,\pk_2,\allowbreak \pi_1,\pi_2)$ where $\pk_1\neq \pk_2$. $\mathcal{A}$ wins if $\Judge(\opk,M,R,\Sigma,\pk_1,\pi_1)=1$ and $\Judge(\opk,M,R,\Sigma,\pk_2,\pi_2)=1$. Note that $\mathcal{A}$ can produce all secret keys. It guarantees that nobody, including a valid user who has $\sk$ and the opener who has $\osk$, can produce such an ambiguous proof.
\end{itemize}

Some studies have employed accountable ring signatures in blockchain systems~\cite{SatoEFO21,FujitaniEO21,EsginSZ22,EsginZSLL19} due to its decentralized structure. 
In addition, Qiao et al.~\cite{QiaoMA22} removed the opening functionality from group signatures to reduce its centralized structure; however, it still has a centralized structure because a group manager issues signing keys to the group members unlike to (accountable) ring signatures. 
Linkable ring signatures~\cite{LiuWW04,BeullensKP20,LuAZ19} have been employed to provide a pseudonymity level privacy protection. As mentioned previously, this is insufficient; thus, anonymity level privacy protection is desirable.  
Connolly et al.~\cite{ConnollyDLP22} proposed a revocable and auditable anonymous credential scheme called Protego, and they considered its application to Hyperledger Fabric. Although it might be employed in the proposed system, Connolly et al. only considered permissioned blockchains. 
Thus, accountable ring signatures are employed in the proposed system. 

\subsection{Layer 2}
\label{L2}

Ethereum is widely recognized as providing high security as a smart contract-enabled blockchain. However, in terms of ensuring security and providing a distributed structure, there is room for improvement regarding scalability. Concretely, the number of transactions that can be processed within a certain time period is less than that of other blockchain systems. Thus, transaction fees (gas costs) will increase when many EOAs want to issue transactions. To solve this problem, we can issue a transaction off-chain, and the proof that the transaction is generated correctly, is only stored on the Ethereum blockchain. In this case, the blockchain and off-chain are referred to as Layer 1 (L1) and Layer 2 (L2), respectively. 
For example, when a Merkle tree is used to generate a hash chain, the proof that the hash chain satisfies the Merkle tree is stored in L1. To reduce the proof size stored in L1, the zk-STARK (zero-knowledge Scalable Transparent ARgument of Knowledge)~\cite{Ben-SassonBHR18} or zk-SNARK (zero-knowledge Succinct Non-interactive ARguments of Knowledge)~\cite{Groth10a,ParnoHG016} are employed. 

zkSync is an Ethereum L2 network technology that supports zk-SNARK. The goal of L2 technologies to provide the same execution environment of Ethereum (i.e., the EVM: Ethereum Virtual Machine), and zkSync nearly supports the same execution environment of Ethereum, i.e., it is EVM-compatible and can run a smart contract programed by Solidity. Thus, we can utilize Ethereum ecosystem tools, and smart contracts on zkSync are described by Solidity. This means that the proposed system can be used directly when Ethereum supports account abstraction. 


\subsection{Account Abstraction}

\begin{figure}[h]
\centering
\includegraphics[clip,width=9cm]{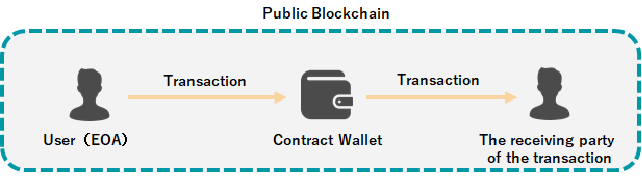}
\vspace{-0.3cm}
\caption{Usual Transaction Issuing}\label{usual}
\end{figure}

\begin{figure}[h]
\centering
\includegraphics[clip,width=9cm]{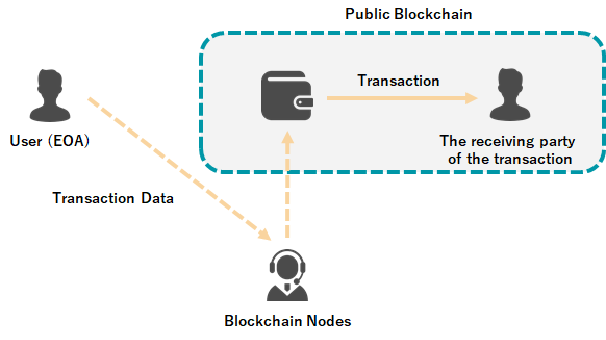}
\vspace{-0.3cm}
\caption{Account Abstraction}\label{fig:acountabstraction}
\end{figure}

In the following, we describe account abstraction. A user sends the code of a contract wallet to nodes that support account abstraction, and requests to its deployment. Note that the user does not need to be an issuer of a transaction. After deploying the contract wallet, an issuer sends transaction data to blockchain via nodes. 
Finally, the contract wallet runs the transaction according to the rule described in the code. We illustrate the flow of transaction issuing when account abstraction is/is not employed in Figure~\ref{usual} and Figure~\ref{fig:acountabstraction}. 

Here, the issuer (User (EOA)) communicates with nodes, their IP address is known by nodes which break anonymity trivially. 
Thus, we exclude the nodes in our anonymity requirement.%
\footnote{This restriction can be removed if a user (who issued a transaction) prepares a node. However, currently only a test net supports account abstraction, and users who can prepare nodes are restricted. } 

\section{Proposed Anonymous yet Accountable Contract Wallet System}

In this section, we describe the proposed anonymous yet accountable contract wallet system. First, we classify users who issue transactions. 

\begin{itemize}
\item Group User: a user $i$ who manages a key pair of the underlying accountable ring signature scheme $(\pk_i,\allowbreak \sk_i)$ and the common opening key $\osk$. $i\in R$ denotes when user $i$ is a group member where $R$ is the ring used for signature verification (i.e., $\pk_i\in R$). 

\item Individual User: a user $j$ who manages own ECDSA key pair $(\vk_j,\allowbreak\sigk_j)$. 
\end{itemize}

\smallskip
\noindent 
For simplicity, we assume that an issuer is either a group user or individual user. Thus, we define a policy $\Policy$ that specifies when the contract wallet runs a transaction. For example, let $\Policy:=\{R,j\}$. Then, a transaction is run when one group user who belongs $R$ and one individual user $j$ agree with the execution of the transaction. In this case, the ring $R=\{\pk_i\}$ and ECDSA verification key $\vk_j$ are registered in the contract wallet. 

\subsection{Security Requirements}

We require the following security notions hold. 
Especially, due to the provability and proving soundness, the protocol is accountable because they allow us to clarify the accountability involved in issuing a transaction. 

\begin{itemize}
\item Anonymity: All entities that can observe transactions including contract wallets and excluding group users and nodes that communicate with the transaction issuer, cannot identify the actual issuer among the verification key holders contained in ring $R$. 

\item Unforgeability: As long as all signatures (either/both the ring signatures and/or the ECDSA signatures) specified by the $\Policy$ are sent to a contract wallet, no entity (who does not have the corresponding signing key) can issue a transaction that is valid under the $\Policy$. 

\item Provability: When a group user $i$ sends a ring signature to issue a transaction, the user can generate its proof. In addition, nobody including group users can issue a transaction where the corresponding user is not identified. 

\item Proving Soundness: When a group user $i$ does not send a ring signature to issue a transaction, the user cannot generate a valid proof that is accepted by the $\Judge$ algorithm with $\pk_i$. 
\end{itemize}

\begin{table*}[t]
\caption{Notations}\label{notations}
\centering
\begin{tabular}{cl} \hline
   \textbf{Notation} & \textbf{Description} \\ \hline\hline
   $(\pk_i,\sk_i)$ & A key pair of the accountable ring signature scheme for a user $i$ \\ 
   $(\opk,\osk)$   & The common opening public and secret keys of the accountable\\
                   &  ring signature scheme \\
   $R$ & A ring (a set of public keys)\\
   $\Sigma$ & A ring signature \\
   $\pi$    & A proof of transaction issuing\\
   $(\vk_j,\allowbreak\sigk_j)$ & A key pair of ECDSA for a user $j$\\
   $\sigma$  & A ECDSA signature \\\hline
\end{tabular}
\end{table*}

\subsection{High-level Description}

In the following, we explain the case of $\Policy:=\{R,j\}$ (See Table~\ref{notations} for notations). 
First, we assume that all group users belonging to $R$ share $\osk$. 
Let a group user $i\in R$ issue a transaction. Then, the user $i$ generates a ring signature $\Sigma$, and the individual user $j$ generates an ECDSA signature $\sigma$, respectively. After these signatures are sent to a contract wallet, the contract wallet checks whether $\Sigma$ is valid under $R$, and $\sigma$ is valid under $\vk_j$. If both signatures are valid, then the contract wallet runs the transaction. We note that only the signature verification is executed on-chain. 
Next, the user $i$ runs the $\Open$ algorithm using $\osk$, generates $\pi$ that is a proof of transaction issuing, and sends $\pi$ to other group users (via an off-chain channel). For example, any information sharing tool used in the organization (e.g., a bulletin board in the organization or e-mails) can be used for sending $\pi$. Other group users can recognize that the user $i$ issues the transaction by checking $\pi$ using the $\Judge$ algorithm. 

We can easily consider a case that consents of two or more individual users are required. For example, $\Policy:=\{R,j,k\}$. Similarly, we can consider two or more rings, e.g., $R_1$ and $R_2$. Here, we assume that $R_1\cap R_2=\emptyset$ and do not consider the case $R_1\cap R_2\neq\emptyset$ because the contract wallet cannot distinguish whether one user who belongs to $R_1\cap R_2\neq\emptyset$ sends two ring signatures or not, due to anonymity. 
Similarly, we do not consider a case that consents two or more group users belonging to the same ring are required because the contract wallet cannot distinguish whether one user who belongs to the ring sends two or more ring signatures or not, due to anonymity. 
We remark that ring members can internally check whether one user who belongs to the ring sends two or more ring signatures or not. Thus, in the case that the proof $\pi$ will be opened after issuing the transaction, these restrictions could be removed. 

\subsection{Proposed System}

\begin{figure}[h]
\centering
\includegraphics[clip,width=7.5cm]{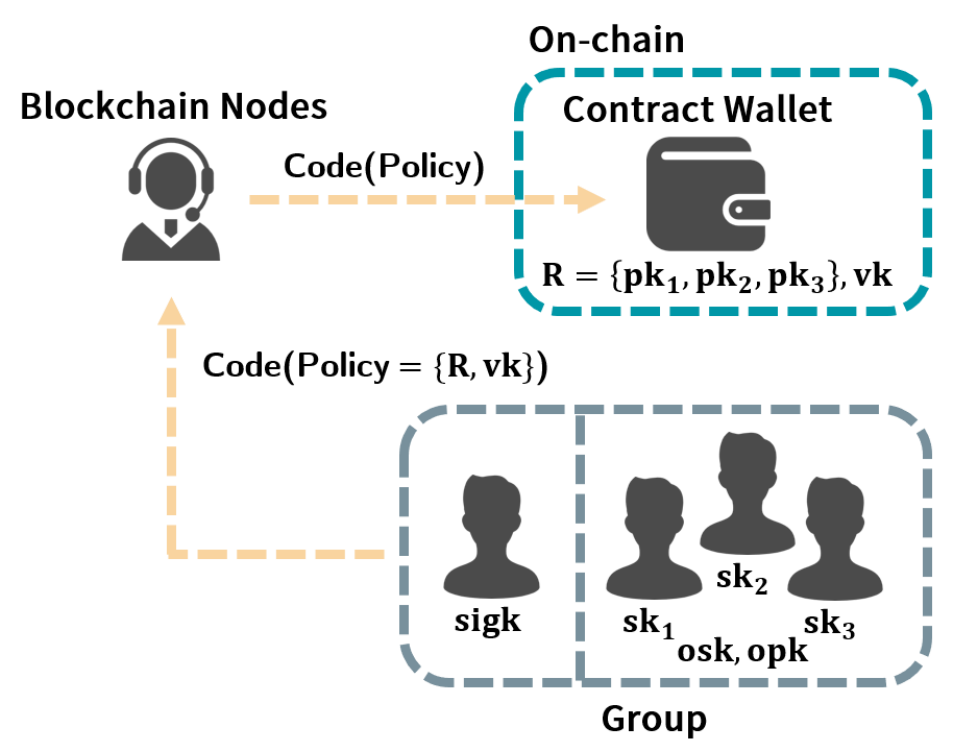}
\vspace{-0.3cm}
\caption{Whole flow of $\DeployContractWallet$}\label{deployContract}
\end{figure}

\begin{figure}[h]
\centering
\includegraphics[clip,width=9cm]{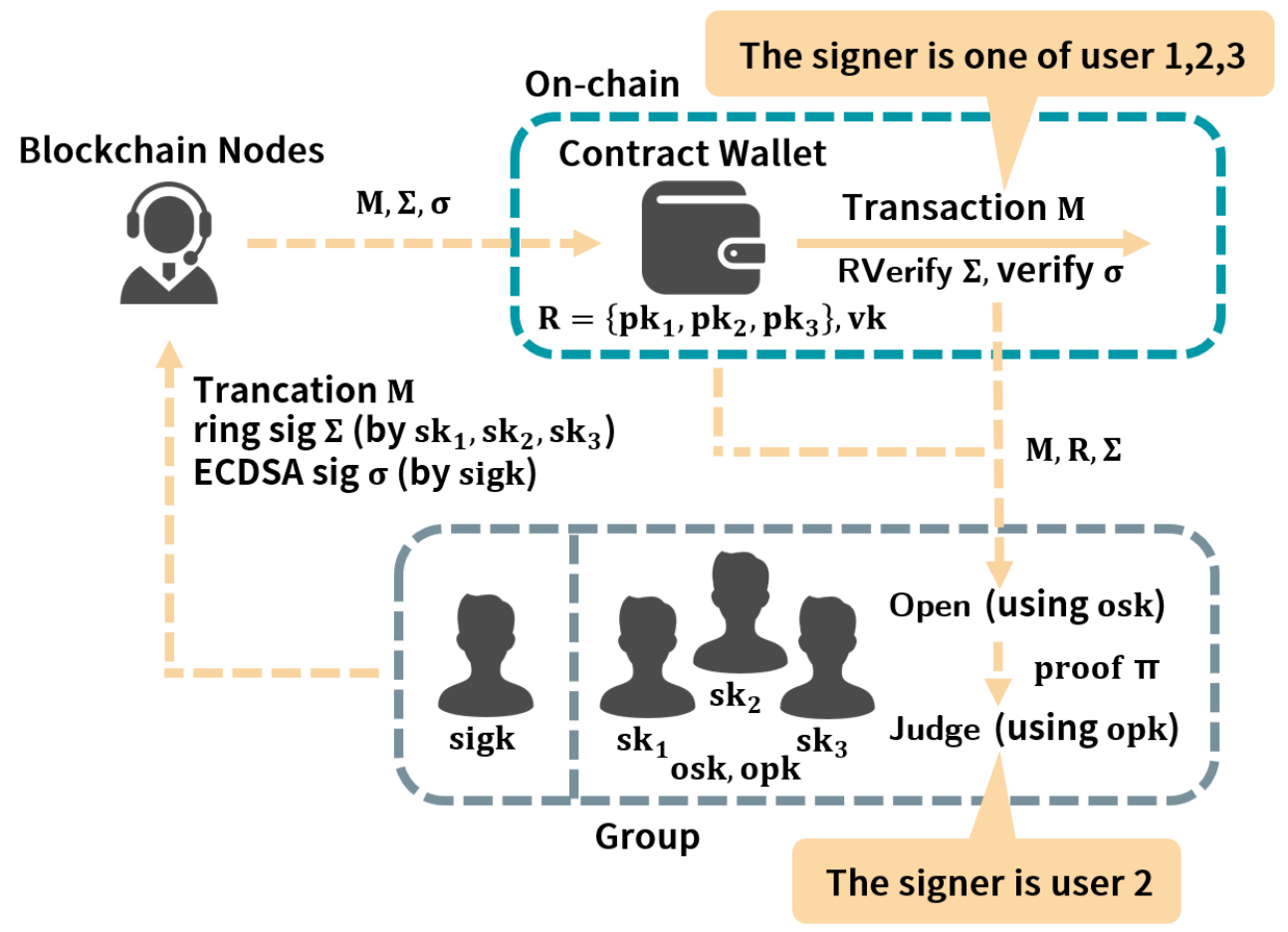}
\vspace{-0.3cm}
\caption{Whole flow of $\SendTransaction$}\label{sendTransaction}
\end{figure}

Here, we describe the proposed anonymous yet accountable contract wallet system. The proposed system consists of two procedures, $\DeployContractWallet$ and $\SendTransaction$. The $\DeployContractWallet$ protocol is used to deploy a contract wallet, and the $\SendTransaction$ protocol issues a transaction against a transaction issuing request. 
We illustrate the whole flow of the two procedures in Figure~\ref{deployContract} and Figure~\ref{sendTransaction}.

\begin{itemize}
\item $\DeployContractWallet(\Code,\BlockChain)$: Let $\Code$ be the code of a contract wallet. In the protocol, $\Code$ is sent to nodes that support account abstraction. Here, we describe $\BlockChain$ as the set of nodes. 
Then, $\Code$ contains a policy $\Policy$, and a set of verification keys (a ring $R$ and ECDSA verification keys), and the rule that specifies the procedure run after the verification of signatures is passed. Finally, deploy a contract wallet $\CW$. 
\smallskip
\item $\SendTransaction(\TransactionData,\BlockChain,\CW)$: A transaction data $\TransactionData$, containing accountable ring signatures and ECDSA signatures, is sent to $\CW$ via nodes described by $\BlockChain$. Then, $\CW$ checks the validity of signatures according to $\Policy$, and issues a transaction. 

\end{itemize}

In our system, 
the $\RVerify$ algorithm is run on-chain by the contract wallet, as in the case of ECDSA signatures, because anyone can issue a transaction if no verification is involved. 
The only information stored on the blockchain is $R$, $\Sigma$, $\vk_j$, and $\sigma$. If the accountable ring signature scheme is secure, no other users can identify the creator of the ring signature, preserving the anonymity of the signer. 

\subsection{Security Discussion}

\begin{itemize}
\item Anonymity: Due to account abstraction and anonymity of the underlying accountable ring signature scheme, the fact that a group user $i$ belonging $R$ generates the signature to issue a transaction is not leaked. More precisely, the underlying accountable ring signature provides anonymity under the full key exposure. That is, anonymity holds against entities that observe transactions even the corresponding signing key is exposed (although we do not require such a stronger-level anonymity). 

\item Unforgeability: Due to full unforgeability of the underlying accountable ring signature scheme and unforgeability of ECDSA, no entity, that does not belong to $R$ or does not have a signing key corresponding to $\vk$ specified by $\Policy$, can issue a transaction. 
Note that full unforgeability guarantees that no adversary can produce a valid ring signature when the adversary does not have the corresponding signing key. 
Due to the tamper resistance of the blockchain, the code of a contract wallet is not modified after its deployment. 
Thus, as long as all signatures (either/both ring signatures and/or the ECDSA signatures) specified by $\Policy$ are sent to a contract wallet, no entity can issue a transaction that is valid under the $\Policy$. 

\item Provability: Let $\Sigma$ be a ring signature generated by $\sk_i$ and verified by a contract wallet. Due to the correctness of the underlying accountable ring signature scheme, $\Judge(\opk,M,R,\Sigma,\pk_i,\pi)=1$ holds. 
Moreover, due to traceability of the underlying accountable ring signature scheme, nobody can produce a valid but untraceable signature even they have all signing keys and $\osk$. 

\item Proving Soundness: Let a ring signature be generated by $\sk$ that corresponds to $\pk$. Due to tracing soundness, nobody can produce $\pi$ that is accepted by the $\Judge$ algorithm with $\pk^\prime\neq \pk$. Note that tracing soundness guarantees that no adversary can produce proofs for a signature that are accepted by the $\Judge$ algorithm for different keys even the adversary can produce  all secret keys. That is, even a group user who has $\sk$ where $\pk\in R$ and $\osk$ cannot produce such an ambiguous proof. 

\end{itemize}

\section{Implementation}

The dominant of the proposed system in terms of cryptographic operations is to verify accountable ring signatures on-chain. Thus, we mainly focus on the accountable ring signature scheme in our implementation. 

We implemented the Bootle et al. accountable ring signature scheme using Node.js. Because it is secure under the Decisional Diffie-Hellman (DDH) assumption, we employed secp256k1 as the underlying elliptic curve, which is known as a DDH-hard curve. We also implemented the $\RVerify$ algorithm using Solidity because the verification procedure is run on-chain (by the contract wallet). 
For ECDSA signature verification, we employed the OpenZeppelin library\footnote{\url{https://docs.openzeppelin.com/}}. 

\subsection{Contract Wallet Deployment}
\label{deploy}

First, we show that gas (which is used in zkSync) for contract wallet deployment (in the case of $|R|=4$ and $|R|=10$). 
zkSync includes a caching mechanism that significantly reduces deployment costs for repeated deployments. 
As of December 8, 2024, zkSync offers three different gas prices(user gas price, fair gas price for computation, fair gas price for pubdata), making it challenging to calculate the precise cost in USD. However, based on the current gas price ranges, the cost of first deployment is estimated to be between 4 USD and 12 USD, while the cost of subsequent deployments is estimated to be between 0.1 USD and 0.3 USD. 

\begin{table}[h]
\centering
\caption{Gas for Contract Wallet Deployment}
\label{solidity_deploy}
\begin{tabular}{l|c|c}
\hline
    & $|R|=4$ & $|R|=10$      \\ \hline
First Deployment    &  22,024,795 {[}gas{]}  & 22,278,676 {[}gas{]} \\ \hline
Subsequent Deployments  &  652,307 {[}gas{]}  & 887,108 {[}gas{]}          \\ \hline
\end{tabular}
\end{table}

\subsection{Implementation by Node.js}

Here, we describe our implementation environment of Node.js in Table~\ref{node_env}. 

\begin{table}[h]
\centering
\caption{Implementation Environment of Node.js}
\label{node_env}
\begin{tabular}{l|l}
\hline
OS  & Ubuntu 22.04.1 LTS \\ \hline
CPU & Intel Xeon E-2288G (3.70 [GHz]) \\ \hline
Memory & 128 {[}GB{]}       \\ \hline
\end{tabular}
\end{table}

We show our implementation results (which represent the average of running times of 100-times executions) of the accountable ring signature scheme in Table~\ref{node_time}. We estimate the case of $|R|=4$ and $|R|=10$. 

\begin{table}[h]
\centering
\caption{Running Times}
\label{node_time}
\begin{tabular}{l|c|c}
\hline
Algorithms     & $|R|=4$ & $|R|=10$   \\ \hline
$\RSign$   & 329 {[}ms{]} &  473 {[}ms{]} \\ 
$\RVerify$   & 276 {[}ms{]} & 387 {[}ms{]}\\ 
$\Open$ & 296 {[}ms{]}  &  393 {[}ms{]}\\ 
$\Judge$ & 287 {[}ms{]}  & 393 {[}ms{]} \\ \hline
\end{tabular}
\end{table}

Although they depend on the ring size, they are reasonable because they are run off-chain, except the $\RVerify$ algorithm in the proposed system. The running time of the $\RVerify$ algorithm here is also important because the $\Open$ and $\Judge$ algorithms internally run the $\RVerify$ algorithm, since the corresponding ring signatures to be opened are required to be valid.

\subsection{Implementation by Solidity}

Here, we describe our implementation environment in Table~\ref{solidity_env}. 

\begin{table}[h!]
\centering
\caption{Implementation Environment}
\label{solidity_env}
\begin{tabular}{l|c}
\hline
Version   & Solidity @0.8.16                      \\ \hline
Compiler  & hardhat-zksync-solc @0.3.9 \\ \hline
Docker Image & matterlabs/local-node:latest2.0                         \\
 & ghcr.io/paradigmxyz/reth:v0.2.0-beta.2                         \\ \hline
\end{tabular}
\end{table}

We show our implementation results (which represent the average of gas of 50-times executions in the case of $|R|=4$ and $|R|=10$) of the accountable ring signature scheme in Table~\ref{solidity_gas}. 
Currently, transaction fee to run zkSync is not strictly established. Therefore, based on the current gas price ranges, the cost for $\RVerify$ is estimated to be between 9 USD and 27 USD.
We note that the pure cost of performing ECDSA verification (excluding other costs such as signature storage cost) is 11,823 gas. 
The main reason of these high costs is inefficiency of elliptic curve operations in Solidity. 
We note that it seems the transaction needs to be divided in the real environment due to the gas limit because there is a limitation on the transaction size that can be executed in a single transaction in the mainnet. 

\begin{table}[h]
\centering
\caption{Gas for Running $\RVerify$}
\label{solidity_gas}
\begin{tabular}{l|c|c}
\hline
Algorithm    & $|R|=4$ & $|R|=10$      \\ \hline
$\RVerify$  & 55,769,832 {[}gas{]}            & 73,757,714 {[}gas{]}          \\\hline
\end{tabular}
\end{table}


\section{Conclusion}

In this paper, we proposed an anonymous yet accountable contract wallet system based on account abstraction and accountable ring signatures. The proposed system is implemented using Solidity for zkSync. 
Moreover, we discussed potential of the proposed system, e.g., medical information sharing and asset management. Since the current implementation results using Solidity show the required costs are expensive, our result here might be regarded as somewhat conceptual. However, to the best of our knowledge, no previous implementation result is known that confirms the cost to run an accountable ring signature scheme in Solidity to date, and we believe that our result can be seen as an important stepping stone to provide anonymity and accountability simultaneously in blockchain systems. 

Investigating other applications of the proposed system will be left to future work. The underlying account ring signature scheme does not provide post-quantum security due to the discrete logarithm-based construction. 
Thus, it is difficult to accept the current construction as a platform to manage large amounts of assets due to the progress of quantum computing. Because a post-quantum accountable ring signature scheme has been proposed in~\cite{BeullensDKLP22,EsginSZ22,EsginZSLL19},%
\footnote{More precisely, a weaker variant of post-quantum accountable ring signature schemes have been proposed in~\cite{EsginSZ22,EsginZSLL19} where no $\Judge$ algorithm is provided. Since our system employs the $\Judge$ algorithm to provide the proving soundness, the $\Judge$ algorithm is mandatory. } it would be interesting to employ the scheme, precisely, how to implement them using Solidity is left to future work.

\medskip
\noindent\textbf{Acknowledgment}: The authors would like to thank Dr. Miyako Ohkubo (NICT) for her invaluable comments and suggestions. This work was supported by JSPS KAKENHI Grant Numbers JP21K11897, JP22H03588, and JP23K24844.


\end{document}